\documentstyle[eqsecnum, aps]{revtex}
\tighten
\begin{document}
\def\<{\langle}
\def\>{\rangle}

\title{ Gribov vs BRST }

\author{F G  Scholtz}

\address{\small{\it Department of Physics, University of Stellenbosch, 7600
Stellenbosch, South Africa}}

\author{S V Shabanov\thanks {An Alexander von Humboldt fellow.
On leave from: Laboratory of Theoretical Physics,
JINR, Dubna, Russia.}} \address{ \small{\it Institute for Theoretical Physics,
FU-Berlin, Arnimallee 14, D-14195, Berlin, Germany}}

\date{\today}

\maketitle

\begin{abstract} We investigate the way in which the Gribov problem is
manifested in the BRST quantization of simple quantum mechanical models by
comparing models with and without a Gribov problem.
We show that the hermiticity and nilpotency of the BRST charge
together with the Batalin-Vilkovisky theorem yield {\em non-trivial}
supplementary conditions on gauge fixing fermions.
If the gauge fixing fermion satisfies the supplementary
conditions, the BRST physical states form a space isomorphic to
the Dirac space, and the BRST formal path integral does not suffer from the
Gribov problem.
The conventional gauge fixing fermion, that
gives rise to the Faddeev-Popov integral,
fails to satisfy the supplementary conditions due to the Gribov problem.
Alternatively, enforcing the conventional gauge fixing fermion, these
supplementary conditions imply restrictions on the BRST physical
states for which the Batalin-Vilkovisky theorem holds.
We find that these BRST physical states  are {\em not}
isomorphic the Dirac states.
This can be
interpreted as a violation of the Batalin-Vilkovisky theorem on the space of Dirac states and implies a
breakdown of unitarity and a general dependence of physical quantities on the
gauge condition.

\end{abstract}

\vspace{0.5cm}

\section{ Introduction.}

The Gribov problem is one of the main obstacles in the non--perturbative
quantization of gauge theories \cite{1}.  Singer's work \cite{2} has shown that
the Gribov problem is deeply rooted in the global structure of the
configuration space of a gauge theory and that it can not be easily avoided
unless one resorts to rather unconventional gauge fixing conditions with mostly
undesirable features.

The way in which the Gribov problem effects the conventional Faddeev--Popov
path integral has been investigated by many authors, and several suggestions as
to how it can be circumvented have been put forward (see e.g. ref. \cite{3} and
references therein; \cite{fujikawa} for some recent developments).
A question which has received much less attention relates
to the effect the Gribov problem has on the BRST quantization.  Fujikawa
suggested that it may be related to a spontaneous breakdown of the BRST
symmetry \cite{4}.  On the other hand Govaerts pointed out that it may be
connected to a violation of the Batalin-Vilkovisky theorem \cite{8}.

Since the discovery of the BRST symmetry of the Faddeev-Popov action,
there existed a believe that the path integral for theories with local
symmetries can be defined as a path integral for an effective theory
with the global BRST symmetry. This is mainly due to the success of
the perturbative Yang-Mills theory. It was pointed out by Neuberger
\cite{neuberger} that this equivalence breaks down beyond the
perturbation
theory. The conventional BRST action may give rise to a zero partition
function due to the Gribov problem as well as to vanishing expectation
values of physical operators.

For some gauge models like topological Yang-Mills theories and
2D topological gravity, the BRST formal path integral can be cured
\cite{topology}. An extension of this approach to realistic theories
is still a problem because the approach  heavily relies on some
features that are inherent to topological gauge theories, although an
interesting proposal has recently been made in \cite{marthin}.

It should be noted that the Hamiltonian path integral allows one
to resolve the Gribov problem \cite{7}. The recipe can be tested
on soluble gauge models \cite{2dym}. The Hamiltonian path integral
can also be given a form that does not depend on any particular
parametrization of the gauge orbit space and, therefore, it is
``coordinate-free''\cite{klsh1}.

One usually wants to maintain an explicit Lorentz invariance of
the path integral. The standard tool to achieve this goal is the
BRST formalism. Due to the Gribov problem even the BRST formal
path integral is ill-defined and has to be modified.
It is important to realise that, regardless
of the Gribov problem, the path integral measure of
a field theory is ill-defined beyond the perturbation theory
(e.g. \cite{zinn}).  Thus it must be regularised before performing any formal manipulations that aim at
the possible cure of the Gribov decease (see, e.g., a lattice approach in \cite{zw}).
Since the BRST formalism has a  well defined operator interpretation,
it seems rather natural to localise the Gribov problem in
the BRST operator approach and amend the path integral
accordingly. This is the aim of this letter.

Our main concern here will therefore be the construction of the physical
subspaces, following the BRST quantization, of models exhibiting the Gribov
problem and those not possessing the Gribov problem.  From this we extract the
distinguishing feature which isolates the effect of the Gribov problem.

We proceed along the conventional lines of BRST quantization and the normal
assumptions of quantum mechanics.  Thus we quantize in the Hilbert space with
conventional inner product and with the assumptions that (1) the BRST operator
is hermitian and (2) physical observables are hermitian on the physical
subspace \cite{5}.

Our approach is somewhat unconventional when we construct the physical
subspace.  As usual we construct the physical states as the zero ghost number
states annihilated by the BRST charge \cite{5}.  In addition, however, we also
require explicitly that physical transition amplitudes are independent from the
gauge fixing fermion, i.e., that the Batalin-Vilkovisky theorem holds \cite{BV}.  This is
done in order to make the tacit assumptions behind the Batalin-Vilkovisky
theorem explicit. In so doing, we end up with some
supplementary conditions
on admissible gauge fixing fermions, which are crucial for the
sequel analysis of the Gribov problem.  These conditions are model independent and apply to any BRST quantized system.
Once  these supplementary conditions have been identified, one can take two points of view.  One can
either consider them as a constructive procedure for obtaining the admissible class of gauge fixing
fermions which yields a physical subspace, identified as usual through the cohomology, isomorphic to the
Dirac space.  Alternatively, given the gauge fixing fermion one can view these as restrictions on the
physical subspace on which the Batalin-Vilkovisky theorem holds.  This subspace may not be isomorphic
to the Dirac space, which would imply that the BRST quantization with this particular gauge fixing
fermion is not equivalent to Dirac quantization and hence it should be discarded.  Stated differently, the
Batalin-Vilkovisky theorem is violated on the full Dirac space for this gauge fixing fermion.  This implies
a breakdown of unitarity and leads to a dependence of physical quantities on the gauge condition in the full
Dirac space.

Here we analyse the Gribov problem for
the conventional BRST quantization using a conventional
gauge fixing fermion which leads to the Fadeev-Popov path integral.
Using this gauge fixing fermion, we find in the presence of the
Gribov problem that no globally well defined gauge fixing condition exists
which yields a physical subspace isomorphic to the set of Dirac states.  In
particular such a gauge fixing condition can not be of the class $C^2$
everywhere.  The class of conventional gauge fixing fermions with globally well defined gauge fixing
conditions should therefore be considered as invalid.  This seems to confirm the conclusions of \cite{8}.

The paper is organised as follows: In section 2 we present the two quantum
mechanical models we study and discuss their BRST quantization.  In section 3
we derive our supplementary conditions on gauge fixing
fermions. Section 4 is devoted to the Gribov problem. We
construct the physical subspaces associated with the models and show how the
Gribov problem manifest itself for the conventional gauge fixing
fermion (the latter fails to  satisfy the supplementary conditions).
In section 5 we present our conclusions and discuss possible ways to
circumvent the Gribov problem in the BRST formalism.

\section{The models}

The Lagrangians describing the models we consider are
\begin{mathletters}
\label{2.1}
\begin{eqnarray}
L_1 &=& \textstyle{\frac{1}{2}} \, (\dot{x}- y \, T \, x)^2 -
V \, (x^2) \; , \label{2.1a}\\ L_2 &= &\textstyle{\frac{1}{2}} \, \dot{x}^2_1 + \textstyle{\frac{1}{2}}
\, (\dot{x}_2 - y)^2 - V \, (x_1) \;,\label{2.1b}
\end{eqnarray}
\end{mathletters}
with $x \in I\!\!R^2$, $y \in I\!\!R$ and $T$ the $2\times2$
anti-symmetric matrix with $T_{12}=1$.  We shall refer to (\ref{2.1a}) as model
I and (\ref{2.1b}) as model II.
Both models exhibit a gauge symmetry.  In the case of model I we have
invariance under the transformation $x \rightarrow e^{\theta T} \, x$, $y
\rightarrow y + \dot{\theta}$ and for model II under $x_2 \rightarrow x_2 +
\theta$, $y \rightarrow y + \dot{\theta}$.  As normal the gauge invariance is
associated with the existence of constraints.
For both models the primary constraint is $p_y=0$ and the secondary
constraint is a generator of gauge transformations in the $x$-space
    \begin{equation}
 \sigma_1 = p_x \, T \, x = 0, \quad \sigma_2 = p_{x_2} = 0 \label{2.2}
     \end{equation}
for models I and II, respectively;
here $p_q$ denotes the momenta canonically conjugate to variable
$q$.  The Hamiltonians associated with the models are ($i=1,\;2$)
\begin{equation}
    \label{2.3}
     H^s_i = \textstyle{\frac{1}{2}} \, p^2_x + V_i \, (x)
\end{equation}
with $V_1$ a function of $x^2$ only and $V_2$ a function of $x_1$ only.

We want to stress that model I has a Gribov problem, but model II not\cite{7}.
Consider for example a gauge fixing condition $\chi \, (x)$ such that it is
continuous and single valued. The equation $\chi \, (x)=0$ determines
a line passing through the origin on the $x$-plane. Due to the
single-valuedness of $\chi(x)$, this line intersects each gauge orbit,
being a circle centred at the origin, at least twice.
For model II it is, however, possible to find gauge
conditions with the above properties such that the gauge condition has a unique
solution (e.g. $\chi(x)=x_2=0$).

To BRST quantize,
one extends the classical phase space to include the Lagrange
multiplier $y$ and its canonically conjugate momenta $p_y$.  One also
introduces ghost and anti--ghost degrees of freedom, associated with the two
constraints in each model, and their canonically conjugate momenta \cite{5}.
The extended phase space is thus ($x, \, p_x, \, y, \, p_y, \, \eta, \,
p_{\eta}, \, \overline{\eta}, \, p_{\overline{\eta}}$) where all variables are
taken to be real. We shall also use the collective notations $q^i,
\theta^{\alpha}$ for boson and ghost coordinates, respectively, and
$p_j, \pi_{\beta}$ for the corresponding momenta.

Quantization proceeds by promoting the canonical variables
to linear operators
acting in some Hilbert space and satisfying the canonical
commutation relations
\begin{equation}
[q^i \, , \; p_j] = i \, \delta^i_j\;,\quad [\theta^{\alpha} \, , \,
\pi_{\beta}] = \delta^{\alpha}_{\beta}\;,\label{2.12}
\end{equation}
where $[A , B] = AB - (-1)^{\epsilon_A \, \epsilon_B} \, BA$ is the
graded commutator ($\epsilon$ is the Grassmann parity, $\epsilon
=0,1 $ for even and odd variables, respectively).
One introduces the ghost number operator  \cite{5}
\begin{equation}
N = \eta \, p_{\eta} - \overline{\eta} \, p_{\overline{\eta}}\ ,
\ \ \ \ \epsilon_N = 0 \; , \label{2.6}
\end{equation}
and the nilpotent BRST charge for the two models \cite{5}

\begin{equation}
      Q_i = p_y \, p_{\overline{\eta}} + \eta \, \sigma_i\ ,
\ \ \ \ Q_i^2 = 0\ ,\ \ \ \ \epsilon_Q = 1 \; . \label{2.7}
\end{equation}
Finally one extends the dynamics to the full phase space by introducing the
BRST extended Hamiltonian \cite{5}
\begin{equation}
H_i = H^s_i + [Q_i \, , \; \rho] \label{2.8}\;.
\end{equation}
Here $\rho$ is a gauge fixing fermion of the general form, i.e.,
a generic operator of the ghost number -1, $[\rho,N]=-\rho$.
Note that
$\eta, \, p_{\overline{\eta}}$ have ghost
number 1, while $\overline{\eta}, \; p_{\eta}$ have ghost number $-1$.  Thus
$Q_i$ have ghost number 1.
Any system
observable can be extended to the full phase space in this way.  One easily
verifies that
\begin{equation}
[Q_i \, , \; H_i] = [N \, , \; H_i]= 0 \; , \label{2.10}
\end{equation}
so that the ghost number and BRST charge are conserved.

We adopt the Schr\"odinger realisation for the operator algebra
(\ref{2.12}), i.e., the Hilbert space is
the space of functions $\psi \, (x, \, y, \,
\overline{\eta}, \, \eta)$ with inner product \cite{5}
\begin{equation}
\< \psi \, | \, \phi \> \;
= \; i \, \textstyle{\int }\, dx \, dy \, d \overline{\eta} \, d
\eta \, \psi^{\ast} \, \phi \label{2.13}
\end{equation}
(note that $( d\bar\eta d\eta)^*=- d\bar\eta d\eta$, therefore the factor $i$).
For the moment we leave the boundary conditions on the function $\psi \, (x, \,
y, \, \overline{\eta}, \, \eta)$ open.  We shall address this issue below.  The
operator algebra (\ref{2.12}) is then realised by
$q^i  \rightarrow  q^i\ ,\ p_i  \rightarrow  - i \,
\partial/\partial q^i$ and for the ghosts
$ \theta^{\alpha}\rightarrow  \theta^{\alpha},\  \pi_{\beta}
\rightarrow  \vec{\partial}/{\partial \theta^{\beta}}$.

The ghost number
provides a grading on the Hilbert space and we can write
\begin{equation}
V = V^0 + V^{-1} + V^1\,, \label{2.15}
\end{equation}
where $V^0$, $V^{-1}$ and $V^1$ are, respectively, the ghost number 0, -1 and 1
states
\begin{equation}
\psi^0 = A \, \overline{\eta} \, \eta + B \, \in \, V^0 \; ,\quad
\psi^{-1} = C \, \overline{\eta} \, \in \, V^{-1}\;,\quad
\psi^1 = D \, \eta \, \in \, V^1\;.\label{2.16}
\end{equation}
As the ghost number operator is anti--hermitian, one easily verifies that
\cite{5}
\begin{equation}
\<\psi^n \, | \, \psi^m\> \;
= \; 0 \quad \mbox {\rm unless} \quad n + m = 0 \; .
\label{2.17}
\end{equation}

The conventional BRST procedure is to identify the physical subspace with the
zero ghost number cohomology of $Q$, i.e.,
\begin{equation}
V_{ph} = H^0 = \frac{({\rm Kern} ~Q)^0}{({\rm Im}~Q)^0} \; . \label{2.18}
\end{equation}
The rational behind this procedure is that $Q-$exact states are orthogonal to
physical states and have zero norm.  Thus they do not contribute when the
matrix elements of any physical observables, i.e., an observable that commutes
with $Q$, are calculated.  Hence $Q-$exact states should be identified with the
null vector, which leads us to the cohomology.  The hermiticity of $Q$ is
rather crucial for this procedure.  The reason is that in the absence of
hermiticity one fails to conclude that for any state $\psi^0$ such that $Q
\psi^0 = 0$, $\<\psi^0 \, | \, Q \, \phi \> = 0$
or that $\<Q \, \phi \, | \, Q \,
\psi \> = 0$. Thus, we have to reduce the Hilbert space $V$ to a
subspace
$V_Q$ such that
\begin{equation}
\< Q\psi |\phi\> = \<\psi |Q\phi\>\ , \ \ \
\< Q\psi |Q\phi\> =0\ ,
\label{vq}
\end{equation}
for any $\psi,\phi \in V_Q\subset V$. The first condition ensures
the hermiticity of $Q$, while the second condition can be regarded as
a consistency condition for the first one, i.e., $QV_Q \subset V_Q$,
which follows from the nilpotency of $Q$.

For a unifying treatment of the two models, let us denote the constraints
$\sigma_j = - i \, \frac{\partial}{\partial q^j}$.
For model I $\sigma_1 = - i
\, \frac{\partial}{\partial \phi}$, $\phi \, \in \, [-\pi, \, \pi)$ and for
model II $\sigma_2 = - i \, \frac{\partial}{\partial x_2}$, $x_2 \, \in \, (-
\infty, \, \infty)$. We also denote $q_0 = \pi, \infty$ for models I
and II, respectively.
By making use of the decomposition (\ref{2.16}) and
the orthogonality relation (\ref{2.17}),
one easily verifies that hermiticity of $Q$ requires

\begin{mathletters}
\label{2.19}
\begin{equation}
\textstyle{\int} \, dx \, A^{\ast} \, C\vert_{y=-\infty}^{y=+\infty}  =
\textstyle{\int} \, d\tilde{x} \, dy \, B^{\ast} \, C
\vert_{q=-q_0}^{q=+q_0} = 0\ ,
\label{2.19a}
\end{equation}
while the second condition in (\ref{vq}) yields
\begin{equation}
\textstyle{\int }\, dx \, (\sigma \, C_2)^{\ast} \, C_1
\vert_{y=-\infty}^{y=+\infty}
 = \textstyle{\int} \, d \tilde{x} \, dy \, (p_y \, C_2)^{\ast}
\, C_1\vert_{q=-q_0}^{q=+q_0}
= 0 \; . \label{2.19b}
\end{equation}
\end{mathletters}
Here $d \tilde{x}$ denotes the measure
after the $q$ degree of freedom has been integrated out.

To simplify formulas, we adopt the notation $F^{\pm y}$
for the asymptotic behaviour of
the function $F$ at $y=\pm \infty$ and similarly $F^{\pm q}$ for
$F$ at $q=\pm q_0$.
From the first condition in (\ref{2.19b}) follows $(\sigma C)^{\pm y}
=0$ for both models so $C^{\pm y}$ is gauge invariant. For the
first relation in (\ref{2.19a}) to hold, we have to require that
$A^{\pm y}$ does not have a gauge invariant part, i.e., $\int_{-q_0}^{q_0}dq A^{\pm y}=0$.  Thus we
have the conditions
\begin{mathletters}
\label{2.20}
\begin{equation}
\sigma C^{\pm y} = 0 \, ,\quad A^{\pm y} = \sigma \bar{a}^{\pm y}\, ,\quad (\bar{a}^{\pm
y})^{+q}=(\bar{a}^{\pm y})^{-q}\,.
\label{2.20a}
\end{equation}
The second conditions in (\ref{2.19b}) gives $p_y C_I^{+q}=p_y C_I^{-q}$ and $p_y C_{II}^{\pm
q}=0$ for models I and II, respectively.  The second condition in (\ref{2.19a}) implies for model I
$B_I^{+q}=B_I^{-q}$, while $\int_{-\infty}^{\infty}dy B^{\pm q}=0$ for both models.  The boundary
conditions are therefore
\begin{eqnarray}
p_yC^{+q}_{I} = p_yC^{-q}_{I} \, &,& \quad p_yC^{\pm q}_{II} = 0\,, \label{2.20b}\\
B^{+q}_{I} = B^{-q}_{I}\, ,\quad B^{\pm q}& =& p_y \bar {b}^{\pm q}\, ,\quad (\bar {b}^{\pm
q})^{+y}=(\bar {b}^{\pm q})^{-y}\, .
\label{2.20c}
\end{eqnarray}
\end{mathletters}
The indices I and II refer to models I and II, respectively, while the index is omitted if it applies to both
models. All other boundary conditions are still open and will be fixed
below.

A point to note is that a sufficient, but not necessary, condition to ensure
hermiticity of $Q$ (see Eqs.  (\ref{2.7}) and (\ref{2.13})) is to
impose hermiticity of $p_y$ and $\sigma$.  We refrain from imposing this stronger condition
for two reasons (1) $p_y$ and $\sigma$ are $Q$-exact and thus not physical and (2) it is easy
to see that this condition leads to more stringent boundary conditions on
$A$ and $B$ than (\ref{2.20}) so that it yields a smaller physical subspace.

\section{The physical subspace}

Now that we have identified the Hilbert space in which the physical subspace
should be sought, i.e., the cohomology problem should be solved, we can proceed
to do so.  We therefore have to solve the conditions for a physical state
\cite{5}
\begin{equation}
N \, \psi_{ph} = Q \, \psi_{ph} = 0 \; , \label{3.1}
\end{equation}
under the  conditions (\ref{2.20}).

Conventionally the physical subspace, $V_{ph}$, is defined as the quotient
space obtained after factoring all $Q$-exact states from the space (\ref{3.1}).
It is then argued that physical transition amplitudes do not dependent on the
gauge fixing fermion, i.e., that the Batalin-Vilkovisky theorem holds.  This
argument is based on the observation that

\begin{equation}
e^{-itH_i}|\psi_{ph}\rangle = e^{-itH^s_i}|\psi_{ph}\rangle +
Q|\phi\rangle\label{3.2}
\end{equation}
where $H_i$ and $H_i^s$ were defined in (\ref{2.8}) and (\ref{2.3}),
respectively.  Since $Q$-exact states are orthogonal to physical states and
have zero norm, the gauge dependence, which only enters through the $Q$-exact
part, drops out from the physical transition amplitude.  Although this argument
is formally correct, it tacitly assumes that the state $|\phi\rangle$ belongs
to the Hilbert space on which $Q$ is hermitian and nilpotent, i.e.,
$\phi \in V_Q$. In particular, when applied to models I and II, the boundary
conditions (\ref{2.20}) should
hold for this state.  Clearly this is by no means guaranteed
and depends on the properties of the gauge fixing fermion because
$\phi$ depends on it.
One can therefore
take the point of view that the gauge fixing fermion has to be restricted to a
certain class. Thus we impose this explicitly: The admissible gauge
fixing fermion is the one for which the Batalin-Vilkovisky theorem
holds and the quotient space $V_{ph}$ is isomorphic to the set
$V_{Dir}$ of the Dirac states.

Let us now turn to derivation of the supplementary conditions on $\rho$.
For a physical transition amplitude to be independent of the gauge fixing
fermion, we have to require
\begin{equation}
\<\psi_{ph} \, | \, e^{- i \, t \, H_i} \, \phi_{ph}\> \;
= \; \<\psi_{ph} \, | \,
e^{-i \, t \, H^s_i} \, \phi_{ph}\> \; . \label{3.3}
\end{equation}
Expanding the exponential on the left hand side of (\ref{3.3}) we note that a
generic term has the form
\begin{equation}
\<\psi_{ph} \, | \, Q \, [\rho \, (Q_{\rho})^{m_1} \,
(H^s_i)^{n_1} \, (Q \, \rho)^{m_2} \, (H^s_i)^{n_2} \, \dots \, (Q \,
\rho)^{m_n}] \, \phi_{ph}\> \; , \label{3.4}
\end{equation}
where we have used (\ref{3.1}) and $[Q \, , \; H^s_i] = 0$.  We note from
(\ref{3.4}) that for (\ref{3.3}) to hold, we must require
\begin{mathletters}
   \label{3.5}
      \begin{eqnarray}
\rho\psi_{ph} \, &\in& \, V_Q^{-1} \quad \mbox{\rm if}\quad \psi_{ph} \; \in \;
V_{ph}\sim V_{Dir}\;;\label{3.5a}\\[3mm] Q \, \rho \, \psi_{ph} & \in &
V_{ph}\quad \mbox{\rm if} \quad \psi_{ph} \; \in \;
V_{ph}\sim V_{Dir}\; ;
\label{3.5b} \\[3mm] H^s_i \, \psi_{ph} & \in & V_{ph}
\quad \mbox{\rm if} \quad \psi_{ph} \; \in \;
V_{ph} \sim V_{Dir}\ . \label{3.5c}
       \end{eqnarray}
\end{mathletters}
The first condition can be viewed as a {\em kinematical} supplementary
condition on $\rho$. It ensures that in $V_Q$
matrix elements of any gauge invariant
operator, extended to the total BRST Hilbert space according to
the rule (\ref{2.8}), do not depend on the extension.
The validity of the chain of equalities
$\<\psi|[Q,\rho]\phi\> = \<\psi|Q\rho\phi\> =
\< Q\psi|\rho \phi\> =0$ for $Q\phi =Q\psi =0$ implies that the operator
$\rho$ does not throw states out of $V_Q$, $\rho V_Q\subset V_Q$.
It is no trivial
condition because the function $C$ is not arbitrary for  states
belonging to $V_Q$ (cf.
(\ref{2.20})).

The second condition is a simple consequence of (\ref{vq}) and
(\ref{3.5a}). Clearly, if $\rho \psi \in V_Q$ for any $\psi\in V_Q$,
then from (\ref{vq}) follows that $Q\rho \psi\in V_Q$.

The third condition will be referred  to as a {\em dynamical}
supplementary condition since it depends on the system Hamiltonian.
Although this condition may seem trivial at first sight because
$[Q, H^s]=0$, it is, in fact, highly non-trivial. An important point
to note is that this condition is not automatically compatible with the
kinematical condition and, in general, gives rise to some further restrictions
on the gauge fixing fermion.

We stress the following facts:

1) The conditions (\ref{vq}) and (\ref{3.5})
are model--independent and apply to any BSRT quantized gauge theory.
They are by no means guaranteed for a given gauge fixing fermion;

2) If one succeeds to find $\rho$ satisfying (\ref{3.5}), the
corresponding BRST path integral will not suffer from the
Gribov problem at all (if such a problem is present). This follows
from the Batalin-Vilkovisky theorem.  One can therefore take the point of view that these conditions
provide a constructive procedure of finding the class of admissible gauge fixing fermions;

3) Alternatively, if the gauge fermion is fixed, the supplementary
conditions we have obtained can be regarded as restrictions on the admissible
Hilbert space where the Batalin-Vilkovisky theorem holds
and the BRST cohomologies are sought.
In the latter case,
the Gribov problem may destroy the isomorphism $V_{ph}\sim V_{Dir}$.
This indeed takes place for the conventional gauge fixing fermion.  Thus, due to the Gribov problem,
the conventional gauge fixing fermion fails to
satisfy the supplementary conditions as we
demonstrate in the next section.

 \section{Gribov problem}

Let us take the conventional gauge fixing fermion
\begin{equation}
\rho = -y p_{\eta} + \chi(x)\bar\eta\ ,
\label{2.9}
\end{equation}
where $\chi(x)$
is the gauge condition.
With this choice one gets the Faddeev-Popov
path integral in the gauge $\dot{y}-\chi(x)=0$ from the formal
BRST path integral by integrating out canonical momenta.

We construct the physical subspace satisfying the conditions
(\ref{3.1}) and (\ref{3.5}), subject to the boundary
conditions (\ref{2.20}).
A general ghost number zero state has the form $A
\, \overline{\eta} \, \eta + B$. The condition $Q \, \psi_{ph} = 0$ yields
\begin{equation}
\psi_{ph} = (\sigma a -
\sigma \, F) \, \overline{\eta} \, \eta + p_y F \; ,
\label{3.7}
\end{equation}
where $p_y \, a = 0$ and from (\ref{2.20})
\begin{mathletters}
\begin{eqnarray}
\label{3.7aa}
(F^{\pm y})^{ +q}- (F^{\pm y})^{ -q}&=&a^{+q}-a^{-q}\,, \label{3.7a}\\ 
(F^{\pm q})^{ +y}- (F^{\pm q})^{ -y}&=&0
\label{3.7b}
\end{eqnarray}
\end{mathletters}
It is important to note the order of limits here, and to keep in mind that the order can in general not be
interchanged, e.g., $(F^{+y})^{+q}\neq (F^{+q})^{+y}$ in general.

Before proceeding, we calculate the norm of the states (\ref{3.7}).
Invoking the boundary condition (\ref{2.20}) and (\ref{3.7b}) we get
       \begin{equation}
\<\psi_{ph} \, | \, \psi_{ph}\> = \textstyle{\int} dx \,\left\{\,\left ((\sigma
a)^{\ast}  F+(\sigma F)^{\ast} a -
(\sigma \, F)^{ \ast} \, F\right)\vert_{y=-\infty}^{y=\infty}\right\}+\textstyle{\int} d\tilde x
dy(p_yF)^{\ast}F\vert^{q_0}_{-q_0}\,.\label{3.8a}\\
\end{equation}

We first analyse model I.  The kinematical supplementary condition (\ref{3.5a}) gives rise to two conditions
on the new $C$ (see (\ref{2.20a}) and (\ref{2.20b})).  Combining these with (\ref{2.20c}), which yields $p_y
F^{+q}=p_y F^{-q}$, and assuming $\chi$ to be single valued result in
\begin{equation}
[-y \,\sigma^2 (a - F) + \sigma\chi p_y F]^{\pm y} = 0\,,\quad (a-F)^{+q}=(a-F)^{-q}\label{3.9}
\end{equation}
From (\ref{3.9}) follows the asymptotic behaviour $(p_yF)^{\pm y}= yb_0
^{\pm}(x)$ and $(\sigma F)^{\pm y}= f^\pm, \ p_yf^\pm =0$.
These conditions are compatible if $\sigma b_0^\pm =0$, i.e.,
$b_0^\pm$ is gauge invariant. Now we require that physical states
have non-zero norm. First consider the last term in (\ref{3.8a}).  From (\ref{2.20c}), (\ref{3.9}) and
(\ref{3.7b}) we note that this term vanishes.  For an arbitrary $\sigma a \neq 0$ the first
term in (\ref{3.8a}) is finite if $F^{\pm y}< \infty$ and, hence,
$b_0^\pm $ must be zero. The condition (\ref{3.9}) with $b_0^\pm=0$
implies that $F^{\pm y}= \gamma_0^\pm +a,\ \sigma\gamma_0^\pm=0$. Substituting  in (\ref{3.8a}) we
find that the term in curly brackets also vanishes.  Hence the states have zero norm.  Thus, for non-zero
norm states $a\equiv 0$. Note that the states with $a\neq 0$
can be written as $Q$-exact states, $QC\bar{\eta}, \ C=F-a$. With $b_0^\pm=0$
(\ref{3.9}) ensures (\ref{2.20a}), $(\sigma C)^{\pm y}=0$.

If $a=0$, the condition $b_0^\pm =0$ can be dropped. The growing
asymptotic of $F^{\pm y}$ provided by the non vanishing $b_0^\pm$ does
not lead to a divergent norm because it is projected out by $\sigma$ (recall
$\sigma b_0^\pm =0$).
The non-zero contribution to the integral
(\ref{3.8a}) comes from $(\sigma F)^{\pm y} $
which is required to be finite. The condition (\ref{3.9}) leads to
\begin{equation}
\sigma(\sigma  F^{\pm y} +  \chi \, b^{\pm }_0)=0\,,\quad F^{+q}=F^{-q}\, . \label{3.10}
\end{equation}
Decomposing $\chi = \chi_0 + \sigma\chi'\ , \sigma\chi_0=0$ and using (\ref{3.7b}),
we get for the set of physical states (\ref{3.1})
\begin{equation}
\psi_{ph} = \{-  \sigma \, F \, \overline{\eta} \, \eta + p_yF\,  : \,
F^{\pm
y} = iy^2b^{\pm }_0/2 - \chi' \, b^{\pm }_0 \, ,\ \sigma b_0^\pm =0\,,
F^{+q} = F^{-q}\,, (F^{+q})^{\pm y}=(F^{-q})^{\mp y}\}\;, \label{3.11}
\end{equation}
where the single valuedness of $\chi^\prime$ was used to note that (\ref{3.7a}) ($a=0$) is satisfied.   The
norm of these states are given by
\begin{equation}
\<\psi_{ph} \, | \,\psi_{ph}\>_{I}= \textstyle{\int} \, dx \,
[(\chi'\sigma\chi') \left[b_0^{+ \, \ast} b_0^{+} -b_0^{- \, \ast}
b_0^{-} \right]\,.\label{3.12}
\end{equation}

The Gribov problem manifests itself already on the {\em
kinematical} level.
Indeed, the global (in the sense that it is
independent of the state vector) term $\int_{-\pi}^{\pi}
d\theta\chi'\sigma\chi'$ factorizes from (\ref{3.12}).
A single valued
$\chi^\prime$ will be periodic in $\theta$ so that this term vanishes and all
physical states have zero norm as a consequence of the kinematical
supplementary condition (\ref{3.5a}).  This is, however,  easily circumvented since, seeing that this is a
global factor, one can still proceed to calculate expectations values of physical observables by simply
regularising this factor in an appropriate way.  The regularisation dependent global factor $\int_{-\pi}^{\pi}
\chi^\prime\sigma\chi^\prime$ then drops out when calculating expectation
values.

Assuming some
regularisation of this global factor, we have the important
property $V_{ph}\sim V_{Dir}$.
This is easy to see.  Decomposing
$p_yF\equiv B = B_0 + \tilde{B}$ with $\sigma \, B_0 =
0$, we note that $B^{\pm y}_0 = y \, b^{\pm y}_0$.  Consider $B' = B'_0 +
\tilde{B}'$ such that $B^{\prime \pm y}_0 = B^{\pm y}_0 = y \, b^{\pm y}_0$.
Then from (\ref{3.10}) $F^{\prime \pm y} = F^{\pm y}$.  Hence $F' = F + \delta
F$ with $\delta \, F^{\pm y} = 0$ and $B' = B + \delta \, B$ with $\delta
B=ip_y\delta F$. Therefore we can express the difference between two such
states as $Q i\delta F\bar\eta $ where $i\delta F\bar\eta\in V_Q^{-1}$.  We
conclude therefore that the states (\ref{3.11}) mod $Q-$exact states are in a
one-- to--one correspondence with the Dirac states.

Using physical states with zero (unregularised) norm to evaluate the partition function will lead to a
vanishing result.  Furthermore the expectation values of physical operators will be ambiguous.  This fact
lies behind the
path integral analysis with the similar conclusion \cite{neuberger}.
Neuberger has also mentioned that multi-valued gauges may resolve
the problem, which indeed seems to follow from (\ref{3.12}).
One should, however, be careful with such a conclusion. When deriving
(\ref{3.11}), we have assumed the wave functions to be single
valued. If a multi-valued gauge is used, the Hamiltonian (\ref{2.8})
is no longer single-valued so are the wave functions.
Therefore the entire
analysis, starting from hermiticity of $Q$, has to be revised.  It is an open question whether hermiticity of
$Q$ can be reconciled with conventional gauge fixing fermions using multi-valued gauges.  If not, the use
of such gauges would be invalidated.

As was pointed out above, the vanishing of the norm is not really problematic and can be avoided quite
easily.  There is a much more serious difficulty, which can not be circumvented, that the BRST formalism
encounters in the presence of a Gribov problem.   This relates to the violation
of the dynamical supplementary condition
(\ref{3.5c}), which implies that the Gribov problem invalidates the
Batalin-Vilkovisky theorem.
The dynamical supplementary condition implies that the state
\begin{equation}
H^s_1\, \psi_{ph} = -  \, \sigma \, H^s_1 \, F \, \overline{\eta} \, \eta +
H^s_1 \, p_yF \, \equiv \, - \, \sigma \, F' \,
\overline{\eta} \, \eta + p_yF'
\label{3.14}
\end{equation}
should belong to (\ref{3.11}).
Clearly
$p_yF^{\prime \pm y}
= y \, H^s_1 \,b^{\pm }_0 \, \equiv \, y \, b^{\prime\, \pm
}_0$.
 Since the gauge invariant part
$\chi_0$ of $\chi$ is not involved in (\ref{3.11}) and, hence,
will not affect the sequel analysis, we shall assume
it to be zero.
The crucial condition is
$\sigma F^{\prime\, \pm y} = -\chi  b^{\prime\, \pm }_0$,
which yields
$
[\chi, \, H^s_1] \, b^{\pm }_0 = 0$.
To maintain the isomorphism with the Dirac states, this has to hold for
all scalar square integrable functions, i.e., functions of $r^2=x_1^2+x_2^2$
alone.  This is only possible if
\begin{equation}
[\chi, \, H^s_1]  = if(x)\sigma_1 \label{3.16a}
\end{equation}
with $f$ real.  Since $\chi$ is real and $H_1^s$ is hermitian, consistency
requires that $f$ is a function of $r$ only.  Calculating the commutator we
have $
[\chi, \, H^s_1] = (\partial_1\chi)\partial_1
+(\partial_2\chi)\partial_2+\nabla^2\chi/2 $
from which we conclude
\begin{equation}
\label{3.16c}
\partial_i\chi=f(x) T_{ij}x_j\;,\quad
\nabla^2\chi=0\;.
\end{equation}
{}From the first equation in (\ref{3.16c}) we
deduce $\nabla^2\chi= -i\sigma f$ so that the second equation implies
$\sigma f=0$, consistent with our earlier observation that $f$ is a function of
$r$ only.  It follows from (\ref{3.16c}) that
\begin{equation}
   \label{3.16d}
[\partial_2,\partial_1]\chi=f' r +2f\ ,
\end{equation}
i.e.,
in general $\partial_2\partial_1\chi\neq\partial_1\partial_2\chi$
almost everywhere so that $\chi$ is  almost nowhere of class $C^2$.
Demanding equality we solve $f (r)=r^{-2}$ and $\chi=-\arctan(x_2/x_1)$, so
that $\chi$ is everywhere of class $C^2$ except at the origin.  Note also that
$\chi$ is not a single valued function of $x$, indeed $\chi$ is just the angle $\theta$.  Going to polar
coordinates, the
laplacian contains $\sigma_1^2$ so that the commutator with $\chi$ is
proportional to $\sigma_1$ as required.  It is also amusing to note that this gauge is free from the Gribov
problem.

We conclude that there is no globally well defined $\chi$ such that
the dynamical supplementary condition (\ref{3.5c}) holds.
Thus for any globally well
defined gauge condition $\chi$
this implies that the set of states for which the
Batalin-Vilkovisky theorem holds, does not coincide with the set (\ref{3.11})
and, therefore, it is {\em not} isomorphic to the set of Dirac states.
Consequently
we have to interpret this as a breakdown of the Batalin-Vilkovisky theorem on
the space (\ref{3.11}) for this particular class of gauge fixing conditions.  Since this class of gauge fixing
fermions leads to the Fadeev-Popov
prescription, it signals a similar problem there.

The above considerations eliminates the class of conventional gauge fixing fermions with globally well
defined gauge conditions as a valid class.  We have seen above that multi-valued gauges are compatible
with the dynamical supplementary condition.  However, it is not clear that these gauges are compatible
with the hermiticity of $Q$, so that this class of gauge fixing fermions may also be invalid.  To resolve
these difficulties one has to seek more general gauge fixing fermions satisfying the kinematical and
dynamical supplementary conditions.

Let us return to model II. Conditions (\ref{3.5a}) and (\ref{2.20}) now lead to

\begin{mathletters}
   \label{3.17}
        \begin{eqnarray}
[-y \, \sigma^2(a -  F) +\sigma \chi p_yF]^{\pm y} &=& 0 \;
,\label{3.17a}\\[3mm] [i\sigma (a - F) +y\sigma p_yF+ \chi p_y^2  F]^{\pm q} &=& 0
\; . \label{3.17b}
\end{eqnarray}
\end{mathletters}
As before the asymptotic behaviour following from (\ref{3.17a}) is $(p_yF)^{\pm y}= yb_0
^{\pm}(x)$, $(\sigma F)^{\pm y}= f^\pm, \ p_yf^\pm =0$ and $\sigma b_0^\pm =0$.

For a finite norm (see (\ref{3.8a})) we must have for $\sigma a \neq 0$ that
$F^{\pm y} < \infty$ and
$p_yF^{\pm q} < \infty$, so that $b_0^{\pm}= (\sigma p_yF)^{\pm q}=0$.   As before $b_0^{\pm}=0$
implies $F^{\pm y}=\gamma_0^{\pm}+a$, $\sigma\gamma_0^\pm=0$. Substituting, the term in curly
brackets in the norm (\ref{3.8a})
vanishes as in model  I.  Applying $p_y$ to (\ref{3.17b}) and using  $(\sigma p_yF)^{\pm q}=0$ gives
$(\chi p_y^3F)^{\pm q}=0$.  Thus (assuming $\chi^{\pm q}\neq 0$)  we conclude that $F^{\pm q}$ is at
most quadratic in $y$.  However, since $F^{\pm y}<\infty$, only the constant term is allowed so that $p_y
F^{\pm q}=0$ and the second term in (\ref{3.8a}) also vanishes, yielding a zero norm.  As before the
resulting state is $Q$-exact.

If $a = 0$ we have for a finite norm $(\sigma \, F)^{\pm
y} < \infty$ and $p_yF^{\pm q} < \infty$.   The conditions (\ref{3.17}) lead to

\begin{equation}
     \sigma(\sigma F^{\pm y}+\chi  b^{\pm y}_0)=0\,, \quad (-i\sigma F + \chi p_y^2  F)^{\pm q} =
0\,,\quad (\chi p_y^3F)^{\pm q}=0\;.\label{3.18}
\end{equation}
Decomposing  $\chi=\chi_0+\sigma\chi^\prime$, $\sigma\chi_0=0$ and using (\ref{3.7b}) we identify the set
of physical states (\ref{3.1}) as

\begin{equation}
\label{3.19}
\psi_{ph} = \{-  \sigma \, F \, \overline{\eta} \, \eta + p_yF \, : \,
F^{\pm
y} = iy^2b^{\pm }_0/2 - \chi' \, b^{\pm }_0\,,\sigma b_0^\pm =0\,,(\sigma F)^{\pm q}=(p_y F)^{\pm
q}=0\}\,,
\end{equation}
where (\ref{3.7a}) ($a=0$) implies the condition
${\chi^\prime}^{+q}={\chi^\prime}^{-q}$ on the gauge  fixing
function.   The rest of the analysis is as for model I.  The set
(\ref{3.19}) mod $Q-$exact
states is in a one--to--one correspondence with the Dirac states and condition
(\ref{3.5c}) is satisfied provided $[H_2^s,\chi]b_0^\pm=0$.  The main difference now resides in the
structure of $\sigma$.  In
model II $\sigma_2 = p_2$, and one easily checks that this is satisfied
for any function of $x_1$ if we choose $\chi = x_2$, which yields $\chi^\prime=x_2^2/2$.  Thus a globally
well
defined gauge fixing function exists such that the Batalin-Vilkovisky theorem
holds on the space (\ref{3.19}) which is isomorphic to the set of Dirac states.
Returning to the norm (\ref{3.8a}) we note that it reduces, as in model I, to (\ref{3.12}) where  the global
term that factorizes for
$\chi=x_2$ is $\int_{-\infty}^\infty dx_2 x_2^3$, which yields after
regularisation a well defined definite norm.

It is interesting to note that if the potential $V(r)$ of model I has a well
developed minimum at $r\neq 0$, and one expands around this minimum, keeping
only leading order terms, then model I reduces 'locally' to model II so that a
gauge condition like $\chi=x_2$ can be used 'locally' in model II.

\section{Discussion and conclusions.}

We have shown that the hermiticity and nilpotency of the BRST charge
together with the Batalin-Vilkovisky theorem yield non-trivial
supplementary conditions on gauge fixing fermions. The supplementary
conditions apply to a general BRST quantized gauge theory and provide
both the kinematical (Hilbert space) and dynamical (S-matrix)
equivalence of the BRST scheme to the Dirac formalism.
The BRST path integral is not affected by the
Gribov problem, provided the gauge fixing fermion satisfies
our supplementary condition.
These observations were our main goal.

Next we have demonstrated with simple examples of gauge models that
the conventional gauge fixing fermion fails to satisfy
the supplementary conditions if the Gribov problem is present.
As a consequence, the evolution operator (\ref{3.2}) throws
the states out of the physical subspace (\ref{3.11}) $V_{ph}\sim V_{Dir}$
to a larger subspace so that the Batalin-Vilkovisky
theorem does not hold. This fact can also be
interpreted as a breakdown of unitarity of the physical S-matrix.

To circumvent the Gribov problem in the BRST formalism, one can begin
to investigate the following possibilities.
Firstly, one can try to find a gauge fixing fermion that fulfils the
supplementary conditions we have found.
A possible candidate is associated with multi-valued gauges, although it is not clear that this is compatible
with the hermiticity of $Q$.
Secondly, one can consider other possible inner products \cite{6} to
weaken the supplementary conditions themselves.
This will of course lead to unconventional path
integrals.  More drastically one can consider dropping the condition of a
hermitian BRST charge.  This will, however, at the very least, require a
drastic revision of the BRST quantization procedure.  The reason is simply that
one looses the property that BRST exact states are orthogonal to physical
states and, particularly, that BRST exact states have zero norm.  Thus the
basic philosophy of the BRST quantization scheme, namely, that BRST exact
states have to be identified with the null vector as they have zero norm, is
invalidated.  Furthermore one cannot argue that BRST exact states will not
contribute to physical transition amplitudes.

\section{Acknowledgements}
Fruitful discussions
with J.R. Klauder, P. Van Nieuvenhiuzen, H.-C. Ren,
M. Schaden, I. Zahed and D. Zwanziger
on various aspects of the Gribov problem
and this work are gratefully acknowledged.

This work was supported by a grant from the
Foundation of Research Development of South Africa.


\begin{thebibliography}\\

\bibitem{1} V.N.\ Gribov, {\it Nucl.\ Phys.}\ {\bf B139} (1978), 1.

\bibitem{2} I.M.\ Singer, {\it Commun.\ Math.\ Phys.}\ {\bf 60} (1978), 7.

\bibitem{3} F.G. Scholtz and G.B.Tupper, {\it Phys. Rev.} {\bf D48} (1993), 1792.

\bibitem{fujikawa}R. Friedberg, T.D. Lee, Y. Pang and H.-C. Ren,
{\it Ann. Phys.} {\bf 246} (1996), 381;\\
 K. Fujikawa, {\it Nucl. Phys.} {\bf B468} (1996), 355.

\bibitem{4} K. Fujikawa, {\it Nucl. Phys.} {\bf B223} (1983), 218.

\bibitem{8} J. Govaerts, {\it Int. J. Mod. Phys.} {\bf A4} (1989), 173; ibid. 4487.

\bibitem{neuberger} H. Neuberger, {\it Phys.Lett.} {\bf B 183} (1987), 337.

\bibitem{topology} H. Kanno, {\it Lett.Math.Phys.} {\bf 19} (1990), 249;\\
C. Becchi and C. Imbimbo, {\it Nucl.Phys.} {\bf B462} (1996), 571;\\
R. Zucchini, "Reducibility and Gribov Problem in Topological
Quantum Field Theory", hep-th/9607003.

\bibitem{marthin} L. Baulieu, A. Rozenberg and M. Schaden,
{\it Phys.Rev.} {\bf D54} (1996), 7825.


\bibitem{7} S.V. Shabanov, {\it Phys.  Lett.}  {\bf B255} (1991), 398;
{\it Mod. Phys. Lett.}  {\bf A6} (1991), 909.

\bibitem{2dym} S.V. Shabanov, {\it Phys.Lett.} {\bf B318} (1993), 323.

\bibitem{klsh1} J.R. Klauder and S.V. Shabanov, "Coordinate-Free
Quantization of First-Class Constrained Systems", hep-th/9608027.


\bibitem{zinn} J. Zinn-Justin, "Quantum Field Theory and
Critical Phenomena" (2nd Edition), Clarendon Press, Oxford, 1993.

\bibitem{zw} D. Zwanziger, "Lattice Coulomb Hamiltonian and
Static Color-Coulomb Field", hep-th/9603203.

\bibitem{5} M. Henneaux and C. Teitelboim, "Quantization of Gauge Systems",
Princeton University Press, Princeton, 1992.

\bibitem{BV} I. A. Batalin and G. A. Vilkovisky, {\it Phys. Lett.} {\bf B69} (1977), 309.

\bibitem{6} R. Marnelius and M. \"Orgen, {\it Nucl. Phys.} {\bf B351} (1991), 474.



\end{thebibliography}
\end{document}